\documentclass[10pt,conference]{IEEEtran}
\IEEEoverridecommandlockouts

\usepackage{cite}
\usepackage{amsmath,amssymb,amsfonts}
\usepackage{algorithmic}
\usepackage{graphicx}
\usepackage{tikz}
\usetikzlibrary{calc}
\usepackage{textcomp}
\usepackage{xcolor}
\usepackage{booktabs}
\usepackage{url}
\usepackage{bm}
\usepackage[hidelinks]{hyperref}
\definecolor{bestgreen}{RGB}{0,110,0}
\newcommand{\best}[1]{{\color{bestgreen}\bm{#1}}}

\def\BibTeX{{\rm B\kern-.05em{\sc i\kern-.025em b}\kern-.08em
    T\kern-.1667em\lower.7ex\hbox{E}\kern-.125emX}}

\begin{document}

\title{A Cold Diffusion Approach for Percussive Dereverberation\\
}

\author{
\IEEEauthorblockN{Dimos Makris}
\IEEEauthorblockA{\textit{Department of Music Technology and Acoustics}\\
\textit{Hellenic Mediterranean University}\\
Rethymno, Greece\\
dimakr@hmu.gr}
\and
\IEEEauthorblockN{András Barják}
\IEEEauthorblockA{\textit{XLN Audio}\\
Stockholm, Sweden\\
andras.barjak@xlnaudio.com}
\and
\IEEEauthorblockN{Maximos Kaliakatsos-Papakostas}
\IEEEauthorblockA{\textit{Department of Music Technology and Acoustics}\\
\textit{Hellenic Mediterranean University}\\
Rethymno, Greece\\
maximoskp@hmu.gr}
}

\maketitle

\begin{abstract}

Most recent advances in audio dereverberation focus almost exclusively on speech, leaving percussive and drum signals largely unexplored despite their importance in music production. Percussive dereverberation poses distinct challenges due to sharp transients and dense temporal structure. In this work, we propose a cold diffusion framework for dereverberating stereo drum stems (downmixes), modeling reverberation as a deterministic degradation process that progressively transforms anechoic signals into reverberant ones. We investigate two reverse-process parameterizations—Direct (next-state) and a $\Delta$-normalized residual (velocity-style) prediction—and implement the framework using both a UNet and a diffusion Transformer backbone. The models are trained and evaluated on curated datasets comprising both acoustic and electronic drum recordings, with reverberation generated using a combination of synthetic and real room impulse responses. Extensive experiments on in-domain and fully out-of-domain test sets demonstrate that the proposed method consistently outperforms strong score-based and conditional diffusion baselines, evaluated using signal-based and perceptual metrics tailored to percussive audio. Code and audio examples are available online\footnote{\href{https://github.com/dimakr169/drums_dereverb}{https://github.com/dimakr169/drums\_dereverb}}.


\end{abstract}


\section{Introduction}\label{sec:introduction}

\IEEEPARstart{A}{udio} enhancement algorithms aim to improve the quality of recorded sound by suppressing unwanted distortions such as background noise or reverberation~\cite{hendriks2013dft}. In particular, reverberation – the persistence of sound due to reflections in an acoustic environment – can smear temporal and spectral details of audio, thereby degrading clarity and intelligibility. Therefore, \emph{dereverberation} is the process of removing or reducing reverberant effects and has long been recognised as a crucial task for improving the overall perceived quality, specifically on speech ~\cite{kinoshita2013reverb}.

Indeed, most prior research in audio enhancement has focused on speech signals, motivated by applications in robust telecommunication, automatic speech recognition, and speaker separation~\cite{wang2018supervised}. Early speech dereverberation methods ranged from classical signal processing techniques (e.g., estimating and subtracting late reflections via a weighted prediction error algorithm~~\cite{nakatani2010speech}) to modern deep learning models that directly learn to map reverberant speech to its anechoic counterpart (e.g.,~\cite{ernst2018speech, wang2020deep}). In parallel, generative approaches, such as diffusion-based, have recently gained traction in this domain with prominent results~\cite{fu2019metricgan, su2020hifi, lu2021study, lu2022conditional}. Within diffusion-based speech dereverberation, score-based formulations in the complex STFT domain have demonstrated strong performance and flexibility~\cite{welker2022speech,richter2023speech}. Despite this ongoing rapid progress~\cite{lemercier2024diffusion}, the methodological and evaluation conventions of the field remain strongly speech-centric. 


In contrast, dereverberation for \emph{music} signals has received comparatively less attention and is complicated by the diversity of sources, production effects, and perceptual objectives~\cite{yasuraoka2010music}. Recent music-oriented dereverberation studies have primarily focused on vocals (often under artificial reverb) and highlight the difficulty of collecting sufficiently diverse paired dry/wet data~\cite{saito2023unsupervised}.
For \emph{percussive} material, the problem is especially delicate: Drum hits have sharp onsets and fast-decaying transients, and reverberation can blur onset timing and redistribute energy into the tail, which impacts rhythmic clarity and downstream analysis/production operations~\cite{wilmering2010effects}.

Reverberation is commonly added to drums in music production for aesthetic effect, but there is growing interest in tools that can remove or reduce reverb from recorded percussion – for instance, to tighten up a drum mix or adapt a recording to a different acoustic context. Moreover, many of the commercial “de-reverb” plugins are typically optimised for dialogue/vocals (e.g., Waves Clarity\footnote{\url{https://www.waves.com/plugins/clarity-vx-dereverb}}) and may not generalise well to percussive inputs. Therefore, this work targets a specific and practically motivated setting: \textbf{Drums Dereverberation}. Unlike isolated single-hit drum samples, stereo downmixed drums exhibit dense transient activity and overlapping resonances across kit elements, which complicates blind dereverberation~\cite{grindlay2008blind}. Additionally, percussion lacks continuous harmonic content or phonetic structure to guide a model; small errors in a drum’s transient or timbre can be perceptually obvious. These factors make percussive dereverberation a distinct and challenging task.

To the best of our knowledge, this is the first attempt to investigate a learning-based solution specifically for blind dereverberation of percussive signals. We adopt and extend \emph{cold diffusion}, a family of diffusion-like iterative restoration methods that replace additive Gaussian corruption with a \emph{deterministic} degradation operator and learn to invert it through iterative reconstruction~\cite{bansal2023cold}. Cold diffusion has recently been instantiated for speech enhancement~\cite{yen2023cold}, demonstrating that deterministic degradation schedules can yield strong restoration performance and robustness. It has also been used in music source separation settings where a structured forward process enables learning from imperfect targets~\cite{plaja2023carnatic}.

Inspired by these developments, we define a deterministic forward process that interpolates between anechoic and reverberant complex stereo spectrograms. We study two reverse prediction modes:
(i) \emph{Direct (next-state) prediction}, which predicts the next less-reverberant intermediate spectrogram at each step, and
(ii) \emph{$\Delta$-normalized residual (velocity-style) prediction}, which predicts the step-size-normalized difference between consecutive intermediate spectrograms.
We instantiate both with a UNet~\cite{song2020score} and a diffusion Transformer~\cite{peebles2023scalable} (DiT), and compare them against two strong diffusion baselines originally developed for speech enhancement/dereverberation: SGMSE+~\cite{welker2022speech,richter2023speech} and CDiffuSE~\cite{lu2022conditional}. 


Since most related diffusion-based  pipelines are designed and evaluated in the speech domain, this is reflected not only in the training data and degradations but also in common evaluation protocols. Widely used perceptual/intelligibility metrics such as PESQ, POLQA, and ESTOI are explicitly designed for speech quality or intelligibility assessment~\cite{rix2001perceptual,ITU_P863,jensen2016algorithm}. More broadly, objective quality measures can be strongly \emph{application-domain dependent} and may not transfer reliably across domains such as speech, music and source separation~\cite{torcoli2021objective}. Consequently, adapting such evaluation practice directly to percussive inputs is not well-motivated. Thus, we propose a standard signal-oriented evaluation scheme with a dedicated set of \emph{percussive-oriented} low-level and high-level objective metrics tailored to transient-rich audio. Experimental results demonstrate that our cold diffusion models consistently outperform these baselines, supporting cold diffusion as an effective and practical paradigm for transient-rich dereverberation.



The remainder of this paper is organised as follows: Sec.~\ref{sec:method} presents the proposed cold diffusion framework and model variants, Sec.~\ref{sec:exp_setup} describes the dataset and experimental setup including baselines, Sec.~\ref{sec:results} reports results and examples, while Sec.~\ref{sec:conclusions} concludes the paper.


\section{Method}\label{sec:method}

\begin{figure}[t]
\centering
\begin{tikzpicture}
  \node[inner sep=0] (img) {\includegraphics[width=\columnwidth]{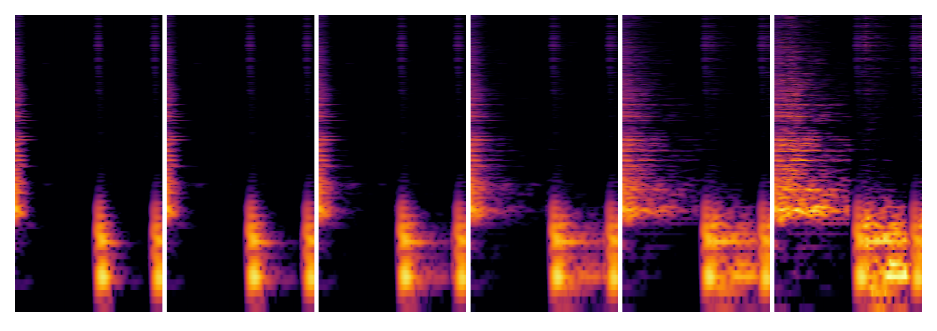}};

  \node[draw,circle,inner sep=1.0pt, font=\scriptsize,
        yshift=0.28cm, xshift=0.25cm] (x0f) at (img.north west) {$x_0$};
  \node[draw,circle,inner sep=1.0pt, font=\scriptsize,
        yshift=0.28cm, xshift=-0.25cm] (xTf) at (img.north east) {$x_T$};

  \draw[->,thick]
    ([xshift=0.12cm]x0f.east) --
    node[above, font=\scriptsize]{Forward process}
    ([xshift=-0.12cm]xTf.west);

  \node[draw,circle,inner sep=1.0pt, font=\scriptsize,
        yshift=-0.28cm, xshift=0.25cm] (x0r) at (img.south west) {$x_0$};
  \node[draw,circle,inner sep=1.0pt, font=\scriptsize,
        yshift=-0.28cm, xshift=-0.25cm] (xTr) at (img.south east) {$x_T$};

  \draw[<-,thick]
    ([xshift=0.12cm]x0r.east) --
    node[below, font=\scriptsize]{Reverse process}
    ([xshift=-0.12cm]xTr.west);

\end{tikzpicture}
\caption{Illustration of the cold diffusion process on a spectrogram. In the forward process, the anechoic signal $\mathbf{x}_0$ is progressively mixed with its reverberant counterpart, while the reverse process learns to iteratively remove reverberation starting from the reverberant signal $\mathbf{x}_T$.}
\label{fig:cold_diffusion_process}
\end{figure}
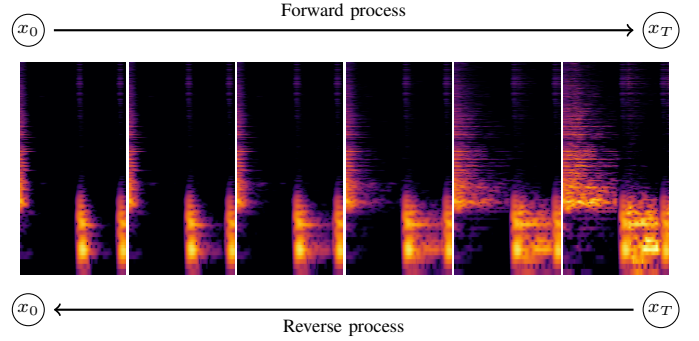

Following recent enhancement work in the complex STFT domain~\cite{welker2022speech,richter2023speech}, we operate on the \emph{stereo} complex spectrogram of the drum stem. Audio is segmented into 2\,s excerpts and transformed using an STFT with FFT size $1024$ and hop size $384$, yielding $F$ frequency bins and $K$ time frames. For each excerpt, we compute complex STFTs for the left and right channels, $X^{(L)},X^{(R)} \in \mathbb{C}^{F \times K}$, and represent them using real and imaginary (RI) components. Concretely, we stack the channels as
$
\mathbf{x} \in \mathbb{R}^{4 \times F \times K}
$
with
$
\mathbf{x} = [\Re\{X^{(L)}\},\Im\{X^{(L)}\},\Re\{X^{(R)}\},\Im\{X^{(R)}\}] .
$
This representation preserves phase information and allows the model to directly learn stereo-consistent dereverberation in the complex domain.

\subsection{Forward Process}
We adopt a \emph{cold diffusion} formulation, where the forward ``process'' is defined via a deterministic degradation operator rather than additive Gaussian noise~\cite{bansal2023cold}. Let $\mathbf{x}_0$ denote the anechoic (clean) stereo RI spectrogram and $\mathbf{y}$ the corresponding reverberant spectrogram. We define a sequence $\{\mathbf{x}_t\}_{t=0}^{T}$ that deterministically interpolates between $\mathbf{x}_0$ and $\mathbf{y}$ using a time-dependent mixing schedule:
\begin{equation}
\mathbf{x}_t \;=\; a_t\,\mathbf{x}_0 \;+\; (1-a_t)\,\mathbf{y},
\qquad a_0=1,\; a_T=0,
\label{eq:forward_linear_interp}
\end{equation}
where $\{a_t\}_{t=0}^{T}$ is a monotone schedule. In our implementation, we use a cosine-squared schedule
\begin{equation}
a_t \;=\; \cos^2\!\Big(\frac{\pi}{2}\frac{t}{T}\Big), \qquad t=0,\dots,T,
\label{eq:alpha_cos2}
\end{equation}
which yields a smooth progression from $a_0=1$ (clean) to $a_T=0$ (fully reverberant). For the $\Delta$-normalized parameterization we further define the schedule step size
\begin{equation}
g_t \;=\; a_{t-1}-a_t, \qquad t=1,\dots,T,
\label{eq:gt_def}
\end{equation}
which is used both to update the state in the reverse process (Eq.~\eqref{eq:reverse_delta_update}) and to form the normalized update target in training (Eq.~\eqref{eq:v_target_simplified}). 

\subsection{Reverse Process}
\label{sec:reverse}
Given $\mathbf{x}_T=\mathbf{y}$, we aim to iteratively recover $\mathbf{x}_0$ by learning a reverse transition model $f_\theta(\cdot,t)$ conditioned on the step index $t$.
We explore two reverse parameterizations:

\vspace{0.2em}
\noindent\textbf{(i) Direct prediction:}
In this mode, the network predicts the next (less reverberant) state:
\begin{equation}
\widehat{\mathbf{x}}_{t-1} \;=\; f_\theta(\mathbf{x}_t, t),
\qquad t=T,\dots,1,
\label{eq:reverse_direct}
\end{equation}
and the reverse trajectory is obtained by iterating Eq.~\eqref{eq:reverse_direct} from $\mathbf{x}_T$ down to $\widehat{\mathbf{x}}_0$.

\vspace{0.2em}
\noindent\textbf{(ii) $\Delta$-normalized residual prediction:}
In this case, the network predicts a step-size-normalized residual increment between consecutive intermediate states:
\begin{equation}
\widehat{\mathbf{v}}_t \;=\; f_\theta(\mathbf{x}_t, t),
\label{eq:reverse_delta_pred}
\end{equation}
which is converted into the next state using the step size $g_t$ from Eq.~\eqref{eq:gt_def}:
\begin{equation}
\widehat{\mathbf{x}}_{t-1} \;=\; \mathbf{x}_t \;+\; g_t\,\widehat{\mathbf{v}}_t.
\label{eq:reverse_delta_update}
\end{equation}
This parameterization predicts a schedule-normalized step update, similar in spirit to step-wise deterministic inversion strategies in diffusion restoration~\cite{bansal2023cold,yen2023cold}. Figure~\ref{fig:cold_diffusion_process} illustrates the proposed deterministic forward degradation and the learned reverse process featuring representative percussive signals (i.e. snare and kick).

\subsection{Training objective}
Training uses paired anechoic/reverberant data $\{(\mathbf{x}_0,\mathbf{y})\}$.
For a randomly sampled step $t \in \{1,\dots,T\}$ we construct $\mathbf{x}_t$ and $\mathbf{x}_{t-1}$ using Eq.~\eqref{eq:forward_linear_interp} and train the network with a weighted sum of a frequency-domain loss and a time-domain loss:
\begin{equation}
\mathcal{L} \;=\; \mathcal{L}_{\text{spec}} \;+\; \lambda_{\text{aud}}\,\mathcal{L}_{\text{aud}}.
\label{eq:loss_total_simplified}
\end{equation}
In \emph{Direct} mode, we supervise the next-state prediction with an $\ell_1$ loss,
\begin{equation}
\mathcal{L}_{\text{spec}}^{\text{direct}}
\;=\;
\left\lVert \widehat{\mathbf{x}}_{t-1} - \mathbf{x}_{t-1} \right\rVert_1.
\label{eq:loss_spec_direct_simplified}
\end{equation}
Regarding \emph{$\Delta$-normalized} mode, we supervise both (i) the reconstructed next state and (ii) a normalized update target. Using Eq.~\eqref{eq:gt_def}, we define
\begin{equation}
\mathbf{v}_t \;=\; \frac{\mathbf{x}_{t-1}-\mathbf{x}_t}{g_t},
\label{eq:v_target_simplified}
\end{equation}
and use a weighted combination of a ``delta'' term and a ``next-state'' term:
\begin{equation}
\mathcal{L}_{\text{spec}}^{\Delta}
\;=\;
0.7\left\lVert \widehat{\mathbf{v}}_{t} - \mathbf{v}_{t} \right\rVert_1
\;+\;
0.3\left\lVert \widehat{\mathbf{x}}_{t-1} - \mathbf{x}_{t-1} \right\rVert_1.
\label{eq:loss_spec_delta_simplified}
\end{equation}
$\mathcal{L}_{\text{aud}}$ is computed after inverse STFT between the estimated and reference waveforms for step $(t-1)$ ($\ell_1$ loss). In practice, we set a larger weight $\lambda_{\text{aud}}=8$ to emphasize waveform fidelity. Finally, in all experiments we use $T=16$ diffusion steps, starting inference from $\mathbf{x}_T=\mathbf{y}$ and iterating the learned reverse transitions to obtain $\widehat{\mathbf{x}}_0$. Preliminary experiments showed no consistent benefit from using more than $16$ reverse steps, so we retained the smallest effective setting.

\subsection{Backbones}

We consider two different backbone network architectures.

\vspace{0.25em}
\noindent\textbf{UNet:} It is based on the NCSN++ architecture commonly used in score-based diffusion models~\cite{song2020score}. We only modify the \emph{input projection} convolution layer with a $(9,1)$ kernel to better match the transient-dominated nature of drum spectra. The network follows a standard encoder--decoder design with four resolution levels and two residual blocks per level, starting from a base width of 64 channels. We include an attention block at the bottleneck to improve global time--frequency context aggregation. This results in a network capacity of roughly 54.6M parameters.

\noindent\textbf{Transformer Diffuser (DiT):} We additionally consider a diffusion transformer that operates on patch-tokenized spectrograms~\cite{peebles2023scalable}. We employ rotary positional embeddings for sequence modeling~\cite{su2024roformer} and inject diffusion-step information through an auxiliary timestep embedding that modulates token representations~\cite{guimaraes2025ditse}. We use a stack of $5$ layer blocks with $8$ attention heads and embedding dimension of $768$, resulting in a 57.1M parameters model, and therefore enables a fair comparison with the UNet backbone.
\section{Experimental Setup}\label{sec:exp_setup}

\subsection{Dataset}

We construct a paired anechoic--reverberant dataset using a combination of the MUSDB18-HQ dataset~\cite{rafii2019musdb18} and the Groove MIDI Dataset (GMD)~\cite{gillick2019learning}. MUSDB18-HQ provides high-quality ($44.1$\,kHz, stereo) music stems from real recordings, from which we extract downmixed drum stems corresponding to physically recorded drum kits. To complement these with electronically generated material, we additionally include drum performances rendered from GMD, yielding a mixed content with diverse rhythmic and spectral characteristics.

Because both source datasets may contain printed reverberation, room ambience, or production effects, all drum stems were manually curated by an experienced audio engineer. Only stems judged to be perceptually \emph{dry}---i.e., lacking audible room tails, algorithmic reverb, or heavy ambience---were retained. This manual filtering is crucial for dereverberation, where the boundary between the ``clean'' signal and reverberant components is inherently ambiguous, particularly for percussive instruments. Drum sounds include resonances from the shell and membrane, sympathetic vibrations across kit elements, and early reflections that may be perceived as part of the source rather than the room. Furthermore, common algorithmic and analog reverbs (e.g., plate or spring simulations) introduce modulation and non-linearities (e.g., gating artifacts) that do not correspond to physical room acoustics and are outside the scope of room dereverberation~\cite{fletcher2012physics}. Our dataset, therefore, targets physically plausible room reverberation, including both early reflections and late reverberant decay, rather than production-style effects.


The selected dry drum mixtures are randomly segmented into 2\,s excerpts.
To increase variability in timbre and production style, we apply data augmentation using the \texttt{audiomentations} library\footnote{\url{https://github.com/iver56/audiomentations}}, including pitch shifting, time stretching, and random equalization. Reverberant counterparts are generated by convolving the dry segments with room impulse responses (RIRs). Following prior work in speech enhancement~\cite{richter2023speech}, we use \texttt{pyroomacoustics}~\cite{scheibler2018pyroomacoustics} to synthesize artificial RIRs with randomized room geometries and reverberation times. To further improve realism and reduce bias toward simulated acoustics, we additionally incorporate measured RIRs from the OpenAIR database~\cite{shelley2010openair}. For each segment, a synthetic or real RIR is randomly selected. Dry excerpts are normalized before RIR rendering; reverberant counterparts then undergo wet/dry energy control and peak safety. Finally, the resulting dataset comprises approximately 38 hours of stereo audio and is available upon request.

\subsection{Baselines}
\label{sec:baselines}

We compare the proposed cold-diffusion models against two representative diffusion-based baselines originally developed for speech enhancement and dereverberation: \textbf{SGMSE+}~\cite{richter2023speech} and \textbf{CDiffuSE}~\cite{lu2022conditional}. To isolate the effect of the diffusion formulation in a controlled comparison, we retrain both baselines using the same data, input representation (stereo RI spectrograms), backbone architecture (UNet), and training configuration as our models. The only difference lies in the diffusion formulation and training objective prescribed by each method. This reduces confounds from backbone capacity, input features, and optimization settings. Inference step counts, however, follow each method’s original configuration and are therefore not matched for computational cost or inference latency: our method uses $T=16$, SGMSE+ is evaluated with 30 reverse steps using the original predictor-corrector sampler with one-step correction, and CDiffuSE uses 50 reverse steps. The comparison should therefore be interpreted as a standard-configuration comparison rather than a matched-compute benchmark.

\subsection{Training configuration}
\label{sec:training}

The dataset is split into 80\% / 10\% / 10\% train, validation, and test sets. All models are trained using the Adam optimizer with a learning rate of $10^{-4}$. During training, we maintain an exponential moving average (EMA) of the model parameters with decay $0.995$; at inference, we use the EMA weights rather than the raw training weights, as this commonly improves stability and perceptual quality in diffusion models~\cite{richter2023speech,lemercier2023storm}.


For the transformer-based diffuser (DiT), we restrict our experiments to the \emph{$\Delta$-normalized} mode. In our preliminary inference experiments, \emph{direct} per-step spectrogram prediction led to unstable reverse diffusion, consistent with the general exposure-bias / error-accumulation phenomenon in diffusion sampling chains~\cite{ning2023input}. The $\Delta$-normalized formulation mitigates this effect by constraining each reverse step to a normalized, schedule-aware update, resulting in stable and reliable inference for the DiT backbone.

\subsection{Evaluation Metrics}


Standard objective measures in dereverberation are largely \emph{speech-oriented} (e.g., PESQ~\cite{rix2001perceptual}, POLQA~\cite{ITU_P863}, (E)STOI~\cite{jensen2016algorithm}) and are not well matched to drum downmix dereverberation. We therefore report two complementary groups of metrics: 

\subsubsection{Low-level (signal-based) metrics}
We report low-level metrics that quantify waveform fidelity, time--frequency magnitude/phase consistency, and improvement over the reverberant input. Let $\mathbf{y}, \mathbf{x}, \hat{\mathbf{x}} \in \mathbb{R}^{2 \times T}$ denote the stereo reverberant input, anechoic reference, and dereverberated estimate, respectively.


\begin{itemize}
\item[(a)] \textbf{Multi-resolution STFT magnitude MAE ($\text{mSTFT}_{\text{mag}}$):}
We measure magnitude fidelity using a multi-resolution STFT log-magnitude distance~\cite{steinmetz2020auraloss}. Multiple resolutions are important for drums because short windows capture sharp transients, while long windows capture decay tails and low-frequency resonances. We average the MAE over $n_{\mathrm{fft}}\in\{256,1024,4096,8192\}$ with hop sizes $\{64,256,1024,2048\}$.

\item[(b)] \textbf{Multi-resolution STFT phase MAE ($\text{mSTFT}_{\text{phase}}$):}
We compute phase error on complex STFTs as the mean absolute wrapped phase difference between $\hat{\mathbf{x}}$ and $\mathbf{x}$ (wrapping to $(-\pi,\pi]$). Phase-aware modeling and evaluation has been shown to impact perceived quality in enhancement tasks~\cite{tan2019learning}.

\item[(c)] \textbf{Error-to-Signal Ratio (ESR):}
ESR~\cite{steinmetz2020auraloss} measures the \emph{residual error energy} relative to the target energy, and is sensitive to gain mismatches:
\begin{equation}
\mathrm{ESR}(\hat{\mathbf{x}},\mathbf{x})=
\frac{\lVert \hat{\mathbf{x}}-\mathbf{x}\rVert_2^2}{\lVert \mathbf{x}\rVert_2^2+\epsilon}.
\end{equation}
The normalization by $\|\mathbf{x}\|_2^2$ enables fair comparison across excerpts with different dynamic ranges.

\item[(d)] \textbf{SI-SDR improvement (SI-SDRi):}
We report \emph{improvement} (SI-SDRi) rather than absolute SI-SDR~\cite{le2019sdr} because dereverberation is a hard condition where absolute SI-SDR can remain low even for perceptually improved outputs (e.g.,~\cite{welker2022speech}). We do not report SI-SIR/SI-SAR, whose interference/artifact decomposition is less meaningful for single-source dereverberation.

\item[(e)] \textbf{Normalized Mutual Information (NMI):}
We compute NMI~\cite{foote1997similarity} between target and estimated magnitude spectrograms:
\begin{equation}
\mathrm{NMI}(U,V)=\frac{I(U;V)}{\sqrt{H(U)\,H(V)}},
\end{equation}
where $I(\cdot;\cdot)$ is mutual information and $H(\cdot)$ entropy.
In our setting, NMI captures non-linear similarity and is less sensitive to exact local time-frequency alignment.
\end{itemize}

\begin{table*}[!htb]
\centering
\scriptsize
\setlength{\tabcolsep}{3.5pt}
\caption{In-domain test set results (mean $\pm$ std) on signal-based and perceptual metrics.}
\label{tab:val_metrics_joint}
\resizebox{\textwidth}{!}{
\begin{tabular}{lccccc@{\hspace{3pt}\vrule\hspace{3pt}}cccc}
\toprule
Model & $\text{mSTFT}_{\text{mag}}$ $\downarrow$ & $\text{mSTFT}_{\text{phase}}$ $\downarrow$ & ESR $\downarrow$ & SI-SDRi $\uparrow$ & NMI $\uparrow$ & MSD $\downarrow$ & ENV $\uparrow$ & TTER $\downarrow$ & ONFi $\uparrow$ \\
\midrule
SGMSE+ & $0.12 \pm 0.52$ & $1.32 \pm 0.29$ & $1.35 \pm 0.75$ & $4.06 \pm 6.32$ & $0.36 \pm 0.18$ & $0.29 \pm 0.09$ & $0.62 \pm 0.31$ & $5.90 \pm 4.51$ & $0.08 \pm 0.17$ \\
CDiffuSE & $0.12 \pm 0.49$ & $1.36 \pm 0.24$ & $1.37 \pm 0.73$ & $2.77 \pm 4.01$ & $0.34 \pm 0.18$ & $0.30 \pm 0.10$ & $0.59 \pm 0.31$ & $6.03 \pm 4.08$ & $0.04 \pm 0.17$ \\
\midrule
Cold UNet \emph{$\Delta$-norm} & $\best{0.08 \pm 0.52}$ & $\best{1.21 \pm 0.36}$ & $\best{0.79 \pm 0.74}$ & $\best{11.09 \pm 10.25}$ & $\best{0.55 \pm 0.16}$ & $\best{0.22 \pm 0.09}$ & $\best{0.92 \pm 0.12}$ & $\best{2.07 \pm 2.13}$ & $\best{0.16 \pm 0.21}$ \\
Cold DiT \emph{$\Delta$-norm} & $0.10 \pm 0.52$ & $1.28 \pm 0.31$ & $1.05 \pm 0.83$ & $7.36 \pm 9.10$ & $0.45 \pm 0.17$ & $0.25 \pm 0.09$ & $0.84 \pm 0.19$ & $3.57 \pm 3.24$ & $0.07 \pm 0.21$ \\
Cold UNet \emph{Direct} & $0.09 \pm 0.52$ & $1.23 \pm 0.35$ & $0.88 \pm 0.79$ & $9.91 \pm 10.13$ & $0.52 \pm 0.17$ & $0.23 \pm 0.09$ & $0.89 \pm 0.16$ & $2.72 \pm 3.05$ & $0.14 \pm 0.21$ \\
\bottomrule
\end{tabular}}
\end{table*}

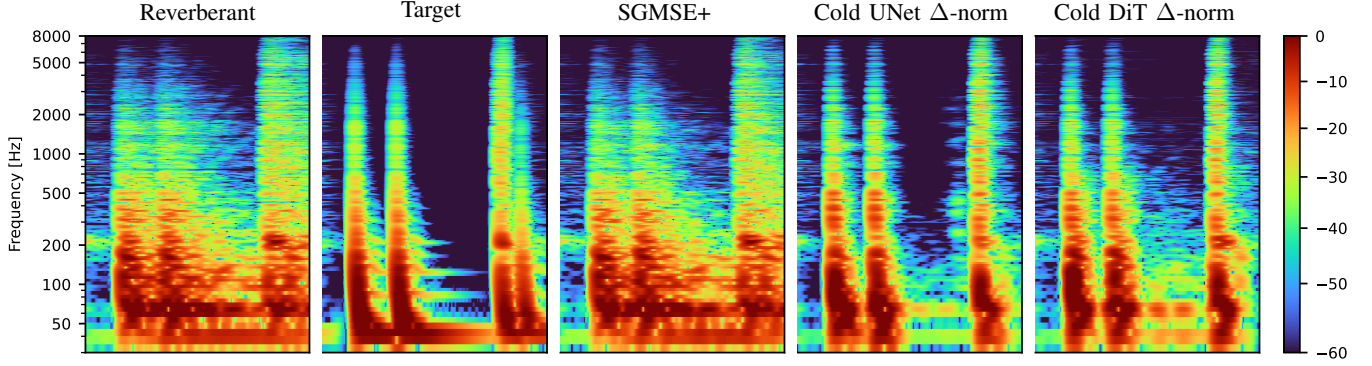
\begin{figure*}[!t] 
\centering
\begin{tikzpicture}
  \node[inner sep=0] (img) {\includegraphics[width=\linewidth]{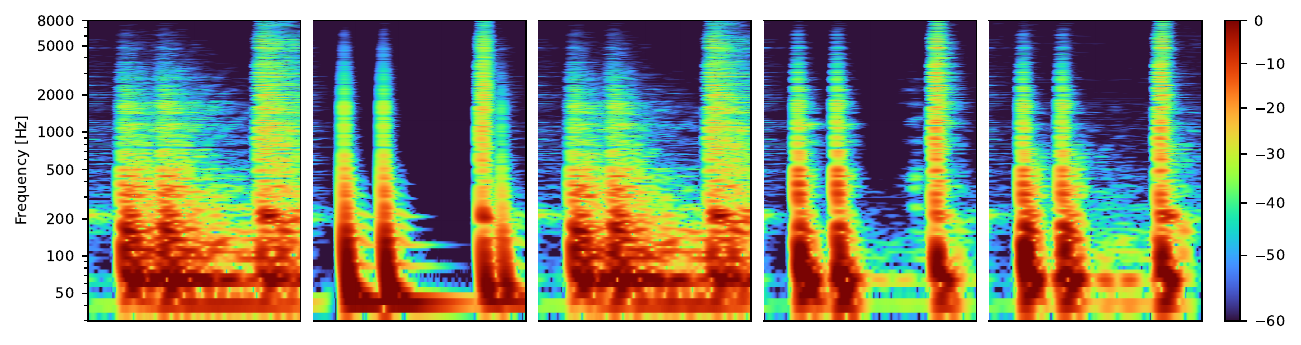}};

  \tikzset{paneltitle/.style={font=\small}} 

  \node[paneltitle, yshift=0.03cm] at ($(img.north west)!0.15!(img.north east)$) {Reverberant};
  \node[paneltitle, yshift=0.03cm] at ($(img.north west)!0.32!(img.north east)$) {Target};
  \node[paneltitle, yshift=0.03cm] at ($(img.north west)!0.49!(img.north east)$) {SGMSE+};
  \node[paneltitle, yshift=0.03cm] at ($(img.north west)!0.67!(img.north east)$) {Cold UNet $\Delta$-norm};
  \node[paneltitle, yshift=0.03cm] at ($(img.north west)!0.84!(img.north east)$) {Cold DiT $\Delta$-norm};

\end{tikzpicture}
\caption{Qualitative spectrogram comparison on an out-of-domain  drum excerpt processed with a highly reverberant unseen impulse response.}
\label{fig:qualitative_ood}
\end{figure*}

\subsubsection{High-level (perceptual) metrics}

Low-level signal-fidelity metrics do not directly quantify perceptually salient aspects of \emph{drums} dereverberation, such as transient clarity, envelope restoration, and preservation of rhythmic event timing. We therefore define the following metrics:


\begin{itemize}
\item[(a)] \textbf{Modulation Spectrum Distance (MSD):}
Reverberation alters temporal-envelope modulations by smearing energy and changing the distribution of modulation frequencies~\cite{falk2010non}. We compute an auditory-inspired modulation representation from subband Hilbert envelopes and measure the distance between target and estimate. Lower values indicate modulation characteristics closer to the anechoic target. 

\item[(b)] \textbf{Envelope Correlation (ENV):}
We compute the Pearson correlation between target and estimated RMS envelopes. This metric is sensitive to transient distortion and tail overhang; higher correlation indicates better preservation of macro-dynamics and event shapes.

\item[(c)] \textbf{Per-hit Transient-to-Tail Energy Ratio deviation (TTER):}
Early-to-late energy balance is related to perceived reverberance~\cite{larsen2008minimum}. For drums, we propose a local onset-conditioned analogue that measures whether dereverberation restores energy concentration near each hit. We detect onset times $\{t_k\}$ on the clean target using \texttt{librosa}~\cite{mcfee2015librosa}.
For each onset, we compute transient and tail energies from fixed windows of length $T_\mathrm{tr}$ and $T_\mathrm{tail}$:
\begin{equation}
R(\mathbf{s};t_k)=10\log_{10}\frac{\sum_{t\in[t_k,\,t_k+T_\mathrm{tr})} s[t]^2 + \epsilon}{\sum_{t\in[t_k+T_\mathrm{tr},\,t_k+T_\mathrm{tr}+T_\mathrm{tail})} s[t]^2 + \epsilon}.
\end{equation}
We report the mean absolute deviation of $R(\hat{\mathbf{x}};t_k)$ from $R(\mathbf{x};t_k)$ across hits.


\item[(d)] \textbf{Onset F-measure improvement (ONFi):}
Dereverberation should sharpen hit boundaries and reduce spurious detections caused by smeared reverberant tails. We therefore compute an onset-detection F-measure by comparing detected onsets against reference onsets obtained from the clean target, using a fixed temporal tolerance window ~\cite{raffel2014mir_eval}. We report an \emph{improvement} score as the difference between the estimate and the reverberant input with respect to the same clean reference. 

\end{itemize}

\section{Results}\label{sec:results}

\subsection{In-domain Evaluation}

Table~\ref{tab:val_metrics_joint} reports in-domain dereverberation performance on the held-out test split. Both speech-oriented diffusion baselines (SGMSE+ and CDiffuse) achieve limited improvement, with similar spectral distortion, high residual energy (ESR), and weak transient–tail separation (TTER), indicating substantial remaining reverberant energy. In contrast, all proposed cold-diffusion models consistently improve both signal-based and perceptual metrics. The best variant, Cold UNet ($\Delta$-norm) achieves the best overall performance, yielding the largest SI-SDR improvement (11.09 dB), the lowest ESR, and the highest envelope correlation, while substantially reducing TTER. These results indicate effective suppression of late reverberation while preserving transient structure. Although the onset F-measure improvement (ONFi) values are numerically modest (best 0.16), this is expected for an improvement in a bounded onset F-measure and still indicates more reliable hit boundary recovery compared to the reverberant input. Finally, $\Delta$-normalized residual inference is consistently stronger than Direct prediction—especially on ESR/TTER/ONFi—supporting the claim that step-normalized updates reduce error accumulation over the iterative reverse process; the DiT backbone benefits similarly but remains below the UNet, particularly on transient-sensitive measures.


\begin{table*}[!htb]
\centering
\scriptsize
\setlength{\tabcolsep}{3.5pt}
\caption{Out-of-domain dataset results (mean $\pm$ std) on signal-based and perceptual metrics.}
\label{tab:test_metrics_joint}
\resizebox{\textwidth}{!}{
\begin{tabular}{lccccc@{\hspace{3pt}\vrule\hspace{3pt}}cccc}
\toprule
Model & $\text{mSTFT}_{\text{mag}}$ $\downarrow$ & $\text{mSTFT}_{\text{phase}}$ $\downarrow$ & ESR $\downarrow$ & SI-SDRi $\uparrow$ & NMI $\uparrow$ & MSD $\downarrow$ & ENV $\uparrow$ & TTER $\downarrow$ & ONFi $\uparrow$ \\
\midrule
SGMSE+ & $0.22 \pm 0.10$ & $1.37 \pm 0.24$ & $1.42 \pm 0.75$ & $2.01 \pm 5.08$ & $0.31 \pm 0.15$ & $0.32 \pm 0.08$ & $0.58 \pm 0.30$ & $6.70 \pm 4.39$ & $0.05 \pm 0.15$ \\
CDiffuse & $0.23 \pm 0.10$ & $1.39 \pm 0.21$ & $1.44 \pm 0.69$ & $0.17 \pm 3.81$ & $0.29 \pm 0.16$ & $0.32 \pm 0.09$ & $0.55 \pm 0.30$ & $6.85 \pm 4.22$ & $0.03 \pm 0.14$ \\
\midrule
Cold UNet \emph{$\Delta$-norm} & $\best{0.16 \pm 0.08}$ & $\best{1.25 \pm 0.33}$ & $\best{1.09 \pm 0.88}$ & $\best{7.52 \pm 8.61}$ & $\best{0.45 \pm 0.17}$ & $\best{0.25 \pm 0.08}$ & $\best{0.84 \pm 0.17}$ & $\best{3.60 \pm 3.35}$ & $\best{0.13 \pm 0.20}$ \\
Cold DiT \emph{$\Delta$-norm} & $0.17 \pm 0.09$ & $1.32 \pm 0.27$ & $1.19 \pm 0.86$ & $5.59 \pm 7.52$ & $0.41 \pm 0.16$ & $0.27 \pm 0.08$ & $0.79 \pm 0.21$ & $4.58 \pm 3.90$ & $0.05 \pm 0.20$ \\
Cold UNet \emph{Direct} & $\best{0.16 \pm 0.08}$ & $1.26 \pm 0.32$ & $1.11 \pm 0.89$ & $7.20 \pm 8.48$ & $0.44 \pm 0.17$ & $\best{0.25 \pm 0.09}$ & $0.83 \pm 0.19$ & $4.00 \pm 3.76$ & $0.11 \pm 0.20$ \\
\bottomrule
\end{tabular}}
\end{table*}

\subsection{Out-of-domain Evaluation}

To evaluate generalization beyond the training distribution, we construct a fully out-of-domain test set using a curated subset of MoisesDB~\cite{pereira2023moisesdb}. Following the same procedure as in the training data, an experienced audio engineer manually selected perceptually dry drum stems, ensuring the absence of existing reverberation.
Reverberant counterparts were generated using \emph{only physical impulse responses}, drawn from real-room recordings (ACE Challenge Workshop\footnote{\url{https://zenodo.org/records/6257551}}). Importantly, this setup introduces a complete domain shift: none of the drum performances, recording characteristics, or reverberation conditions overlap with the training data.

Results are reported in Table~\ref{tab:test_metrics_joint}. As expected, all models exhibit degraded performance compared to the in-domain evaluation, reflecting the difficulty of generalizing dereverberation across unseen drum sources and room acoustics. This degradation is most pronounced for the speech-oriented diffusion baselines, whose SI-SDRi drops to 2.01\,dB (SGMSE+) and near zero for CDiffuSE, alongside increased spectral distortion, residual energy (ESR), and transient–tail imbalance (TTER). These results indicate that baseline diffusion objectives fail to generalize when both the source characteristics and reverberation statistics change.

In contrast, the proposed cold-diffusion models retain substantially better performance across all metric groups. The Cold UNet ($\Delta$-norm) remains the strongest model, achieving more than 7.5\,dB SI-SDR improvement and consistently lower ESR and TTER compared to all baselines. Perceptual proxies confirm this trend: envelope correlation remains high (ENV 0.84), and modulation distortion (MSD) is significantly reduced relative to baseline methods. These results suggest that the deterministic cold-diffusion formulation provides increased robustness to unseen reverberation conditions, likely because the model learns to invert a structured degradation process.

Compared to the in-domain setting, the performance gap between $\Delta$-normalized and direct prediction for the UNet narrows considerably. While the $\Delta$-normalized variant still performs best overall, both modes yield similar scores across most metrics. This behaviour may suggest that under strong domain shift, modeling capacity and inductive bias dominate over fine-grained reverse-process parameterization, and that the UNet backbone itself is the primary driver of robustness. The DiT again benefits from the $\Delta$-normalized formulation but underperforms the UNet, particularly on transient-sensitive metrics (TTER, ONFi).

Figure~\ref{fig:qualitative_ood} shows a qualitative comparison on an out-of-domain electronic drum excerpt processed with a highly reverberant impulse response ($T_{60} > 2\,\mathrm{s}$). The reverberant input exhibits strong energy smearing across time and a pronounced low-frequency bed that masks transient structure. SGMSE+ shows almost no improvement, leaving substantial residual energy and blurred transients. In contrast, both proposed cold-diffusion models substantially reduce late reverberation while preserving sharp onsets; UNet yields the cleanest decay and most compact transient structure. Still, this example also highlights a representative failure case under unseen conditions: UNet partially attenuates low-frequency content, while DiT leaves more residual tail energy. These artifacts suggest that highly reverberant unseen conditions remain challenging despite the overall gains.


\section{Conclusions}\label{sec:conclusions}

We introduced a cold-diffusion framework for stereo drum-stem dereverberation, addressing a largely unexplored percussive enhancement task beyond the speech-centric focus of prior work. By modeling reverberation as a deterministic degradation and evaluating both Direct and $\Delta$-normalized residual reverse parameterizations, our approach consistently outperformed strong diffusion-based baselines on both in-domain and out-of-domain datasets, with improved transient preservation and reduced late reverberation. Future work will focus on expanding the availability of clean percussive data and extending the framework to better handle production-style reverbs, whose non-physical characteristics differ substantially from impulse-response-based room acoustics. We also plan to explore more flexible reverse-diffusion settings, including alternative schedules and variable numbers of reverse steps, to improve robustness and reduce dependence on a fixed inference configuration. Additionally, we will explore using selected metrics from our evaluation suite as potential candidates for auxiliary training losses, aiming to better align optimization with perceptual and transient-sensitive quality.

\section*{Acknowledgment}
This publication is financed by the Project ``Strengthening and optimizing the operation of MODY services and academic and research units of the Hellenic Mediterranean University'', funded by the Public Investment Program of the Greek Ministry of Education and Religious Affairs.

The authors gratefully acknowledge XLN Audio for early support and for discussions that helped shape the initial development and motivation of this work.


\bibliographystyle{IEEEtran}
\bibliography{references}

@article{welker2022speech,
  title={Speech enhancement with score-based generative models in the complex STFT domain},
  author={Welker, Simon and Richter, Julius and Gerkmann, Timo},
  journal={arXiv preprint arXiv:2203.17004},
  year={2022}
}

@inproceedings{rix2001perceptual,
  title={Perceptual evaluation of speech quality (PESQ)-a new method for speech quality assessment of telephone networks and codecs},
  author={Rix, Antony W and Beerends, John G and Hollier, Michael P and Hekstra, Andries P},
  booktitle={2001 IEEE international conference on acoustics, speech, and signal processing. Proceedings (Cat. No. 01CH37221)},
  volume={2},
  pages={749--752},
  year={2001},
  organization={IEEE}
}

@techreport{ITU_P863,
  author      = {{ITU-T}},
  title       = {ITU-T Rec. P.863, ``Perceptual objective listening quality prediction,''},
  institution = {Int. Telecom. Union (ITU)},
  year        = {2018},
  note        = {[Online]. Available: https://www.itu.int/rec/T-REC-P.863-201803-I/en},
  url         = {https://www.itu.int/rec/T-REC-P.863-201803-I/en}
}

@article{jensen2016algorithm,
  title={An algorithm for predicting the intelligibility of speech masked by modulated noise maskers},
  author={Jensen, Jesper and Taal, Cees H},
  journal={IEEE/ACM Transactions on Audio, Speech, and Language Processing},
  volume={24},
  number={11},
  pages={2009--2022},
  year={2016},
  publisher={IEEE}
}

@inproceedings{steinmetz2020auraloss,
  title={auraloss: Audio focused loss functions in PyTorch},
  author={Steinmetz, Christian J and Reiss, Joshua D},
  booktitle={Digital music research network one-day workshop (DMRN+ 15)},
  pages={124},
  year={2020}
}

@article{tan2019learning,
  title={Learning complex spectral mapping with gated convolutional recurrent networks for monaural speech enhancement},
  author={Tan, Ke and Wang, DeLiang},
  journal={IEEE/ACM Transactions on Audio, Speech, and Language Processing},
  volume={28},
  pages={380--390},
  year={2019},
  publisher={IEEE}
}

@inproceedings{le2019sdr,
  title={SDR--half-baked or well done?},
  author={Le Roux, Jonathan and Wisdom, Scott and Erdogan, Hakan and Hershey, John R},
  booktitle={ICASSP 2019-2019 IEEE International Conference on Acoustics, Speech and Signal Processing (ICASSP)},
  pages={626--630},
  year={2019},
  organization={IEEE}
}

@inproceedings{foote1997similarity,
  title={A similarity measure for automatic audio classification},
  author={Foote, Jonathan},
  booktitle={Proc. AAAI 1997 Spring Symposium on Intelligent Integration and Use of Text, Image, Video, and Audio Corpora},
  volume={3},
  year={1997}
}

@book{hendriks2013dft,
  title={DFT-domain Based Single-microphone Noise Reduction for Speech Enhancement: A Survey of the State-of-the-art},
  author={Hendriks, Richard C and Gerkmann, Timo and Jensen, Jesper},
  volume={11},
  year={2013},
  publisher={Morgan \& Claypool Publishers}
}

@article{nakatani2010speech,
  title={Speech dereverberation based on variance-normalized delayed linear prediction},
  author={Nakatani, Tomohiro and Yoshioka, Takuya and Kinoshita, Keisuke and Miyoshi, Masato and Juang, Biing-Hwang},
  journal={IEEE Transactions on Audio, Speech, and Language Processing},
  volume={18},
  number={7},
  pages={1717--1731},
  year={2010},
  publisher={IEEE}
}

@inproceedings{kinoshita2013reverb,
  title={The REVERB challenge: A common evaluation framework for dereverberation and recognition of reverberant speech},
  author={Kinoshita, Keisuke and Delcroix, Marc and Yoshioka, Takuya and Nakatani, Tomohiro and Habets, Emanuel and Haeb-Umbach, Reinhold and Leutnant, Volker and Sehr, Armin and Kellermann, Walter and Maas, Roland and others},
  booktitle={2013 IEEE Workshop on Applications of Signal Processing to Audio and Acoustics},
  pages={1--4},
  year={2013},
  organization={IEEE}
}

@inproceedings{ernst2018speech,
  title={Speech dereverberation using fully convolutional networks},
  author={Ernst, Ori and Chazan, Shlomo E and Gannot, Sharon and Goldberger, Jacob},
  booktitle={2018 26th European Signal Processing Conference (EUSIPCO)},
  pages={390--394},
  year={2018},
  organization={IEEE}
}

@article{wang2020deep,
  title={Deep learning based target cancellation for speech dereverberation},
  author={Wang, Zhong-Qiu and Wang, DeLiang},
  journal={IEEE/ACM transactions on audio, speech, and language processing},
  volume={28},
  pages={941--950},
  year={2020},
  publisher={IEEE}
}

@article{wang2018supervised,
  title={Supervised speech separation based on deep learning: An overview},
  author={Wang, DeLiang and Chen, Jitong},
  journal={IEEE/ACM transactions on audio, speech, and language processing},
  volume={26},
  number={10},
  pages={1702--1726},
  year={2018},
  publisher={IEEE}
}

@inproceedings{fu2019metricgan,
  title={Metricgan: Generative adversarial networks based black-box metric scores optimization for speech enhancement},
  author={Fu, Szu-Wei and Liao, Chien-Feng and Tsao, Yu and Lin, Shou-De},
  booktitle={International Conference on Machine Learning},
  pages={2031--2041},
  year={2019},
  organization={PmLR}
}

@article{su2020hifi,
  title={HiFi-GAN: High-fidelity denoising and dereverberation based on speech deep features in adversarial networks},
  author={Su, Jiaqi and Jin, Zeyu and Finkelstein, Adam},
  journal={arXiv preprint arXiv:2006.05694},
  year={2020}
}

@inproceedings{lu2021study,
  title={A study on speech enhancement based on diffusion probabilistic model},
  author={Lu, Yen-Ju and Tsao, Yu and Watanabe, Shinji},
  booktitle={2021 Asia-Pacific Signal and Information Processing Association Annual Summit and Conference (APSIPA ASC)},
  pages={659--666},
  year={2021},
  organization={IEEE}
}

@inproceedings{lu2022conditional,
  title={Conditional diffusion probabilistic model for speech enhancement},
  author={Lu, Yen-Ju and Wang, Zhong-Qiu and Watanabe, Shinji and Richard, Alexander and Yu, Cheng and Tsao, Yu},
  booktitle={ICASSP 2022-2022 IEEE International Conference on Acoustics, Speech and Signal Processing (ICASSP)},
  pages={7402--7406},
  year={2022},
  organization={Ieee}
}

@article{richter2023speech,
  title={Speech enhancement and dereverberation with diffusion-based generative models},
  author={Richter, Julius and Welker, Simon and Lemercier, Jean-Marie and Lay, Bunlong and Gerkmann, Timo},
  journal={IEEE/ACM Transactions on Audio, Speech, and Language Processing},
  volume={31},
  pages={2351--2364},
  year={2023},
  publisher={IEEE}
}

@article{lemercier2023storm,
  title={Storm: A diffusion-based stochastic regeneration model for speech enhancement and dereverberation},
  author={Lemercier, Jean-Marie and Richter, Julius and Welker, Simon and Gerkmann, Timo},
  journal={IEEE/ACM Transactions on Audio, Speech, and Language Processing},
  volume={31},
  pages={2724--2737},
  year={2023},
  publisher={IEEE}
}

@article{lemercier2024diffusion,
  title={Diffusion models for audio restoration},
  author={Lemercier, Jean-Marie and Richter, Julius and Welker, Simon and Moliner, Eloi and V{\"a}lim{\"a}ki, Vesa and Gerkmann, Timo},
  journal={arXiv preprint arXiv:2402.09821},
  year={2024}
}

@inproceedings{yasuraoka2010music,
  title={Music dereverberation using harmonic structure source model and wiener filter},
  author={Yasuraoka, Naoki and Yoshioka, Takuya and Nakatani, Tomohiro and Nakamura, Atsushi and Okuno, Hiroshi G},
  booktitle={2010 IEEE International Conference on Acoustics, Speech and Signal Processing},
  pages={53--56},
  year={2010},
  organization={IEEE}
}

@inproceedings{saito2023unsupervised,
  title={Unsupervised vocal dereverberation with diffusion-based generative models},
  author={Saito, Koichi and Murata, Naoki and Uesaka, Toshimitsu and Lai, Chieh-Hsin and Takida, Yuhta and Fukui, Takao and Mitsufuji, Yuki},
  booktitle={ICASSP 2023-2023 IEEE International Conference on Acoustics, Speech and Signal Processing (ICASSP)},
  pages={1--5},
  year={2023},
  organization={IEEE}
}

@inproceedings{wilmering2010effects,
  title={The effects of reverberation on onset detection tasks},
  author={Wilmering, Thomas and Fazekas, Gy{\"o}rgy and Sandler, Mark},
  booktitle={Audio Engineering Society Convention 128},
  year={2010},
  organization={Audio Engineering Society}
}

@article{grindlay2008blind,
  title={Blind Dereverberation of Audio Signals},
  author={Grindlay, Graham},
  journal={E4810 Final Project, University of Columbia},
  year={2008}
}

@article{bansal2023cold,
  title={Cold diffusion: Inverting arbitrary image transforms without noise},
  author={Bansal, Arpit and Borgnia, Eitan and Chu, Hong-Min and Li, Jie and Kazemi, Hamid and Huang, Furong and Goldblum, Micah and Geiping, Jonas and Goldstein, Tom},
  journal={Advances in Neural Information Processing Systems},
  volume={36},
  pages={41259--41282},
  year={2023}
}

@inproceedings{yen2023cold,
  title={Cold diffusion for speech enhancement},
  author={Yen, Hao and Germain, Fran{\c{c}}ois G and Wichern, Gordon and Le Roux, Jonathan},
  booktitle={ICASSP 2023-2023 IEEE International Conference on Acoustics, Speech and Signal Processing (ICASSP)},
  pages={1--5},
  year={2023},
  organization={IEEE}
}

@article{plaja2023carnatic,
  title={Carnatic singing voice separation using cold diffusion on training data with bleeding},
  author={Plaja-Roglans, Gen{\'\i}s and Miron, Marius and Shankar, Adithi and Serra, Xavier},
  year={2023}
}

@article{song2020score,
  title={Score-based generative modeling through stochastic differential equations},
  author={Song, Yang and Sohl-Dickstein, Jascha and Kingma, Diederik P and Kumar, Abhishek and Ermon, Stefano and Poole, Ben},
  journal={arXiv preprint arXiv:2011.13456},
  year={2020}
}

@inproceedings{peebles2023scalable,
  title={Scalable diffusion models with transformers},
  author={Peebles, William and Xie, Saining},
  booktitle={Proceedings of the IEEE/CVF international conference on computer vision},
  pages={4195--4205},
  year={2023}
}

@article{guimaraes2025ditse,
  title={DiTSE: High-fidelity generative speech enhancement via latent diffusion transformers},
  author={Guimar{\~a}es, Heitor R and Su, Jiaqi and Kumar, Rithesh and Falk, Tiago H and Jin, Zeyu},
  journal={arXiv preprint arXiv:2504.09381},
  year={2025}
}

@article{su2024roformer,
  title={Roformer: Enhanced transformer with rotary position embedding},
  author={Su, Jianlin and Ahmed, Murtadha and Lu, Yu and Pan, Shengfeng and Bo, Wen and Liu, Yunfeng},
  journal={Neurocomputing},
  volume={568},
  pages={127063},
  year={2024},
  publisher={Elsevier}
}

@article{falk2010non,
  title={A non-intrusive quality and intelligibility measure of reverberant and dereverberated speech},
  author={Falk, Tiago H and Zheng, Chenxi and Chan, Wai-Yip},
  journal={IEEE Transactions on Audio, Speech, and Language Processing},
  volume={18},
  number={7},
  pages={1766--1774},
  year={2010},
  publisher={IEEE}
}

@article{larsen2008minimum,
  title={On the minimum audible difference in direct-to-reverberant energy ratio},
  author={Larsen, Erik and Iyer, Nandini and Lansing, Charissa R and Feng, Albert S},
  journal={The Journal of the Acoustical Society of America},
  volume={124},
  number={1},
  pages={450--461},
  year={2008},
  publisher={AIP Publishing}
}

@article{mcfee2015librosa,
  title={librosa: Audio and music signal analysis in python.},
  author={McFee, Brian and Raffel, Colin and Liang, Dawen and Ellis, Daniel PW and McVicar, Matt and Battenberg, Eric and Nieto, Oriol},
  journal={SciPy},
  volume={2015},
  pages={18--24},
  year={2015}
}

@article{rafii2019musdb18,
  title={MUSDB18-HQ-an uncompressed version of MUSDB18},
  author={Rafii, Zafar and Liutkus, Antoine and St{\"o}ter, Fabian-Robert and Mimilakis, Stylianos Ioannis and Bittner, Rachel},
  journal={(No Title)},
  year={2019},
  publisher={Zenodo}
}

@inproceedings{gillick2019learning,
  title={Learning to groove with inverse sequence transformations},
  author={Gillick, Jon and Roberts, Adam and Engel, Jesse and Eck, Douglas and Bamman, David},
  booktitle={International conference on machine learning},
  pages={2269--2279},
  year={2019},
  organization={PMLR}
}

@book{fletcher2012physics,
  title={The physics of musical instruments},
  author={Fletcher, Neville H and Rossing, Thomas D},
  year={2012},
  publisher={Springer Science \& Business Media}
}

@inproceedings{scheibler2018pyroomacoustics,
  title={Pyroomacoustics: A python package for audio room simulation and array processing algorithms},
  author={Scheibler, Robin and Bezzam, Eric and Dokmani{\'c}, Ivan},
  booktitle={2018 IEEE international conference on acoustics, speech and signal processing (ICASSP)},
  pages={351--355},
  year={2018},
  organization={IEEE}
}

@inproceedings{shelley2010openair,
  title={Openair: An interactive auralization web resource and database},
  author={Shelley, Simon and Murphy, Damian T},
  booktitle={129th Audio Engineering Society Convention 2010},
  pages={1270--1278},
  year={2010}
}

@article{ning2023input,
  title={Input perturbation reduces exposure bias in diffusion models},
  author={Ning, Mang and Sangineto, Enver and Porrello, Angelo and Calderara, Simone and Cucchiara, Rita},
  journal={arXiv preprint arXiv:2301.11706},
  year={2023}
}

@article{pereira2023moisesdb,
  title={Moisesdb: A dataset for source separation beyond 4-stems},
  author={Pereira, Igor and Ara{\'u}jo, Felipe and Korzeniowski, Filip and Vogl, Richard},
  journal={arXiv preprint arXiv:2307.15913},
  year={2023}
}

@article{torcoli2021objective,
  title={Objective measures of perceptual audio quality reviewed: An evaluation of their application domain dependence},
  author={Torcoli, Matteo and Kastner, Thorsten and Herre, J{\"u}rgen},
  journal={IEEE/ACM Transactions on Audio, Speech, and Language Processing},
  volume={29},
  pages={1530--1541},
  year={2021},
  publisher={IEEE}
}

@inproceedings{raffel2014mir_eval,
  title={MIR\_EVAL: A Transparent Implementation of Common MIR Metrics.},
  author={Raffel, Colin and McFee, Brian and Humphrey, Eric J and Salamon, Justin and Nieto, Oriol and Liang, Dawen and Ellis, Daniel PW and Raffel, C Colin},
  booktitle={ISMIR},
  volume={10},
  pages={2014},
  year={2014}
}

\end{document}